# Understanding the network formation pattern for better link prediction

Jiating Yu[1,2] and Ling-Yun Wu[1,2,*]

[1]*IAM, MADIS, NCMIS, Academy of Mathematics and Systems Science,*
*Chinese Academy of Sciences, Beijing 100190, China*
[2]*School of Mathematical Sciences, University of Chinese Academy of Sciences,*
*Beijing 100049, China*

As a classical problem in the field of complex networks, link prediction has attracted much attention from researchers, which is of great significance to help us understand the evolution and dynamic development mechanisms of networks. Although various network type-specific algorithms have been proposed to tackle the link prediction problem, most of them suppose that the network structure is dominated by the Triadic Closure Principle. We still lack an adaptive and comprehensive understanding of network formation patterns for predicting potential links. In addition, it is valuable to investigate how network local information can be better utilized. To this end, we proposed a novel method named Link prediction using **M**ultiple **O**rder **L**ocal **I**nformation (MOLI) that exploits the local information from the neighbors of different distances, with parameters that can be a prior-driven based on prior knowledge, or data-driven by solving an optimization problem on observed networks. MOLI defined a local network diffusion process via random walks on the graph, resulting in better use of network information. We show that MOLI outperforms the other 11 widely used link prediction algorithms on 11 different types of simulated and real-world networks. We also conclude that there are different patterns of local information utilization for different networks, including social networks, communication networks, biological networks, etc. In particular, the classical common neighbor-based algorithm is not as adaptable to all social networks as it is perceived to be; instead, some of the social networks obey the Quadrilateral Closure Principle which preferentially connects paths of length three.

## I. Backgrounds

Networks are widely used in different fields for their powerful ability to model entity relationships that are constantly changing over time, resulting in networks being highly dynamic and complex [1]. The link prediction problem refers to using the information of the observed network, like the properties of the nodes and edges, to predict the existence of edges between pairs of unconnected nodes [2][3]. For static networks, these missing edges may indicate missing information. When we utilize networks to approximately model some complicated systems, for example, missing or redundant edges will inevitably exist due to the complexity of the real systems. For dynamic networks, the structures and properties of the network may be dynamically changing, and some potential edges may appear only in the future. Biological network databases, for example, are constantly updated and enriched as human cognitive capacities improve. Therefore, studying these missing edges is an entry point for studying the link prediction problem, which is crucial for understanding the network formation mechanism.

Most link prediction methods are developed based on common neighbors, with the underlying assumption that the network structure is dominated by the Triadic Closure Principle (TCP): friends of mine are more likely to be friends of each other [4]. However, while this deeply rooted principle is interpretable in some social networks, it's not always reasonable in other networks. For example, Kovács *et al*. pointed out that in Protein-Proteins Interaction (PPI) networks, the link prediction task should rely on paths of length three (L3) rather than length two, because similar proteins tend to recognize the same binding sites [5]. This breaks the long-standing trust of using information of common neighbors. These two opposing ideas are shown in Fig. 1(a), with nodes connected by paths of length two for TCP-based methods and paths of length three for L3 method, and we refer to these two scenarios as the second-order and the third-order information dominated networks, respectively.

However, not all connectivity mechanisms of networks have been researched as much as social networks or PPI networks. When given an arbitrary network whose nodes and edges are of unknown significance, can we make the data tell us whether the second-order neighbors information or the third-order neighbors information is more reliable? Furthermore, even if we know what network information to use, there are still plenty of different ways to exploit them. For example, at least a dozen link prediction algorithms based on common neighbors have been proposed [1]. How can we make better use of the network information to predict potential edges?

---

* lywu@amss.ac.cn

To address these two issues, we developed a novel method named Link prediction using **M**ultiple **O**rder **L**ocal **I**nformation (MOLI), which has two major advantages: first, it enables the diffusion of graphs via the diffusion of random walks on the graph, resulting in better use of network information; second, unlike other prior-driven approaches, it utilizes multiple order neighbors information instead of predetermining whether the second-order or the third-order neighbors information should be used for link prediction. We parameterize the importance of information of different orders, which can be given empirically or obtained by exploring the network evolutionary pattern adaptively. Specifically, MOLI consists of two steps, as shown in Fig. 1(b): first, exploring the pattern of network edge formation by solving an optimization problem or using prior knowledge; and second, applying the results from the first step to predict potential edges via local network diffusion defined on random walks. In detail, given four different network snapshots with weight matrices $W_{old}$, $W_{new}$, $W_{obs}$, $W_{pre}$ of a dynamically changing system, where $W_{old}$ and $W_{new}$ are used to explore the network evolution patterns, and $W_{obs}$ and $W_{pre}$ are used to evaluate the link prediction performance, the workflow of MOLI is as follows:

1. Taking $W_{old}$ and $W_{new}$ as input, we obtain the optimal order coefficient $x$ by solving the optimization problem, whose components correspond to the importance of different orders in interpreting the generation of $W_{new}$ from $W_{old}$. The goal of the optimization problem is to make the edges observed by $W_{new}$ have the highest likelihood of occurrence, while the unobserved edges have the lowest.

2. Taking the order coefficient $x$ obtained in the first step and the observed network $W_{obs}$ ($W_{obs}$ could be $W_{new}$) as input, we generate the output network via local network diffusion, which can be used for the link prediction task with the weights of edges being treated directly as the edge confidence score. The link prediction performance can be evaluated by calculating the area under the precision-recall curve (AUPR) and the area under the receiver operator curve (AUROC) scores with $W_{pre}$.

Note that step 1 can be replaced by prior knowledge if there are no temporal networks available to analyze network evolution pattern, or if we know what information should be used.

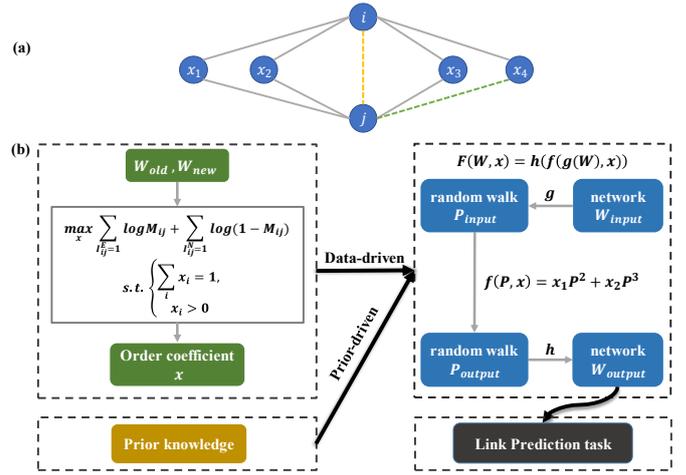

FIG. 1. **(a)** The difference between TCP-based link prediction methods and the L3 method. With grey edges connected, TCP-based methods will connect edge $i-j$, while L3 will connect edge $x_4-j$. **(b)** The pipeline of the MOLI method. We first get the optimal order coefficient $x$ by solving the optimization problem, then take it as input parameter to predict missing links. The output network is obtained based on three operators: the operator $g$ that changes the input graph into the random walk on the graph, and the operator $f$ that defines a diffusion process of random walk, then the operator $h$ that recovers the underlying graph of this diffused random walk [6]. In conclusion, this composite operator changes the input graph into another and adjusts its edge weights. The output network of MOLI can be used for link prediction with the edge's weights being treated directly as the confidence score.

By comparing the link prediction performance of MOLI with 11 other commonly used algorithms on four simulated networks and seven real-world temporal networks, we demonstrate that the proposed method is superior. We are interested in whether there are different patterns in the utilization of local network information for different types of networks. Interestingly, some of the results given by MOLI are consistent with empirical facts, for example, the third-order information is dominant in PPI networks as proposed by Kovács *et al.* [5]. However, we also got some conclusions that contradict the empirical ones. For example, we showed that the second- and third-order local information plays an equally important role in predicting edges of communication networks, while some social networks that empirically obeyed TCP actually confirm to the Quadrilateral Closure Principle (QCP), which preferentially connects paths of length three. These results suggest that prior knowledge is not always trustworthy, and that experience does not always transfer across networks of the similar type. Many network-specific and empirical-based methods are only applicable to some particular networks, while our method can adaptively predict potential edges for various types of networks based on their own formation patterns.

## II. Model

### A. Problem characterization

We formally characterize the link prediction problem as follows: given two networks $G_{obs} = (V, E_{obs})$ and $G_{pre} = (V, E_{pre})$ whose weight matrices are denoted as $W_{obs}$ and $W_{pre}$, where $G_{pre}$ is the network with some new edges added to $G_{obs}$, namely, $E_{obs} \subset E_{pre}$. We need to predict the existence of newly emerging edges $E_{pre} - E_{obs}$ using only the information of $G_{obs}$.

We give some definitions that will be used later. Given two networks $G_{old} = (V, E_{old})$ and $G_{new} = (V, E_{new})$ with $E_{old} \subset E_{new}$ whose weight matrices are denoted as $W_{old}$ and $W_{new}$. $I^E$ is the indicator matrix of the edges emerging in the new network but not in the old network; $I^N$ is the indicator matrix of the edges that have not appeared in either the old or new networks; $I = I^E + I^N$ is the indicator matrix of missing edges in the old network. Mathematically, we have:

$$I^E(i,j) = \begin{cases} 1 & W_{old}(i,j) = 0 \ \& \ W_{new}(i,j) \neq 0 \\ 0 & else \end{cases} \quad (1)$$

$$I^N(i,j) = \begin{cases} 1 & W_{old}(i,j) = 0 \ \& \ W_{new}(i,j) = 0 \\ 0 & else \end{cases} \quad (2)$$

$$I(i,j) = \begin{cases} 1 & W_{old}(i,j) = 0 \\ 0 & else \end{cases} \quad (3)$$

To understand these three matrices more intuitively, we have explained them graphically with an example in Fig. 2.

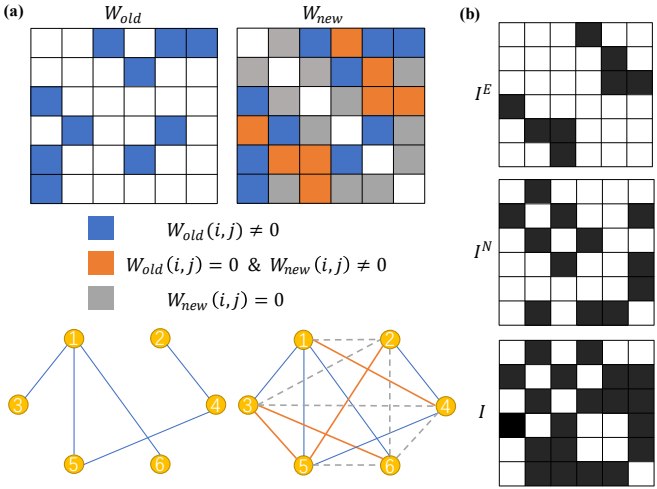

FIG. 2. Schematic of three indicator matrices $I^E$, $I^N$, $I$. **(a)** Given two networks with edge weight matrices $W_{old}$ and $W_{new}$, we use blue to mark those edges in the old network, orange to represent edges that have just emerged in the new network, and gray to indicate edges that still have not appeared in the new network. **(b)** The indicator matrices $I^E$, $I^N$, $I$ correspond to the network given in (a), with black for 1 and white for 0.

### B. Link prediction using Multiple Order Local Information

Here, we elaborate the Link prediction using **M**ultiple **O**rder **L**ocal **I**nformation (MOLI) method, which predicts the presence of possible edges via network diffusion. Unlike other diffusion approaches, MOLI is performed on the transition probability matrix rather than on the (weighted) adjacency matrix of the network directly, following the idea of our previous work [6].

Specifically, three mathematical operators are involved: the operator $f$ which related to the local diffusion process of random walk on the graph, and the operators $g$ and $h$ which help us map this diffusion to the underlying graph. The composite operator of these three operators realizes the diffusion on the graph via the diffusion on the random walk on graph.

In detail, for any input network with weight matrix $W$, let $d$ be the degree vector of input network, that is, $d_i = \sum_j W_{ij}$ for node $i$, and $diag(x)$ be the diagonal matrix whose diagonal elements are vector $x$. Then the operator $g$ defines a random walk on the graph whose (weighted) adjacency matrix is $W$:

$$g(W) = diag(d)^{-1} W \quad (4)$$

It's easy to prove that $g(W)$ is a transition probability matrix.

The operator $h$ has the opposite effect of $g$, which recovers the underlying graph of the random walk defined by the transition matrix $P$:

$$h(P) = diag(\pi(P)) P \quad (5)$$

where $\pi(P) = (\pi_1, \dots, \pi_n)$ is the stationary distribution of transition matrix $P$ such that $\pi P = \pi$, which describes the degree information of underlying graph for random walk defined by $P$.

We have also proved that the operators $g$ and $h$ are inverse operators to each other, so they establish a one-to-one correspondence between graph and random walk on graph (ignoring constant multipliers) [6].

Besides, the operator $f$ defines a local diffusion process on transition probability matrix, which can be mathematically illustrated as:

$$f(P, x) = x_1 P^2 + x_2 P^3 \quad (6)$$

where the components of order coefficient $x = (x_1, x_2)$ represent the importance of the second-order and the third-order diffusion matrices of random walk defined by $P$. Notice that we let $x_1 + x_2 = 1$ and $x_1 \geq 0$, $x_2 \geq 0$ so that the $f(P, x)$ will still be a transition probability matrix.

Collectively, the network operator of MOLI is defined as the composite graph operator $F$ consisting of $g$, $f$ and $h$:

$$F(W, x) = h\big(f(g(W), x)\big) = h(x_1 g(W)^2 + x_2 g(W)^3) \quad (7)$$

which defines the local diffusion on network via the diffusion on random walk.

Notice that the stationary distribution of transition probability matrix keeps unchanged under the operator $f$. Because, assuming that $\pi$ is the stationary distribution of $P$, we have $\pi f(P, x) = \pi(x_1 P^2 + x_2 P^3) = x_1 \pi P^2 + x_2 \pi P^3 = (x_1 + x_2)\pi = \pi$. So, the equation (7) can be rewritten into (denote $diag(d)$ as $D$):

$$F(W, x) = diag(\pi(D^{-1}W))(x_1(D^{-1}W)^2 + x_2(D^{-1}W)^3) \quad (8)$$

And it's easy to check that $dD^{-1}W = \mathbf{1}W = d$, where $d$ is the row vector of degree and $\mathbf{1}$ is the row vector whose elements are all 1. If we normalize $d$ such that $\sum_i d_i = 1$, it is the stationary distribution of $D^{-1}W$. To further simplify the formula (8), we denote $P = D^{-1}W$ and end up with a straightforward and simplified form of MOLI (ignoring constant multipliers):

$$F(W, x) = D(x_1(P)^2 + x_2(P)^3) \quad (9)$$

From equation (9), we can also see that the diffusion matrix of each order of MOLI is actually the product of edge weights on the path normalized by the node degree on the path. Because:

$$\begin{aligned} F(W, x) &= x_1 D(D^{-1}W)^2 + x_2 D(D^{-1}W)^3 \\ &= x_1 W D^{-1} W + x_2 W D^{-1} W D^{-1} W \end{aligned} \quad (10)$$

**C. Learning parameter for MOLI**

The parameter $x$ of MOLI can either be prior-driven or data-driven. When given another two temporal network snapshots $W_{old}$ and $W_{new}$ which come from the same dynamically changing system as the network of interest, we can learn the optimal data-driven parameter $x_{opt}$ for MOLI by solving an optimization problem, which explores the network formation pattern by analyzing how $W_{old}$ generates $W_{new}$.

Denote $M^0$ as the weight matrix generated by $W_{old}$ according to the diffusion model in equation (9):

$$M^0(x) = D(x_1 P_{old}^2 + x_2 P_{old}^3) * I \quad (11)$$

where the $*$ implies the Hadamard (element-wise) matrix product and $I$ is the indicator matrix of missing edges of $W_{old}$, as defined in equation (3). Note that we only consider those missing edges of $W_{old}$, which is the key emphasis for the link prediction task.

Since the values of the $M^0(x)$ directly affect the results of the following optimization problem, we normalize them to eliminate the effect of the magnitude of the matrix elements:

$$M(x) = \frac{1}{\beta(x)} M^0(x), \text{ where } \beta(x) = \sum_{i,j} M_{ij}^0(x) \quad (12)$$

To obtain the optimal parameter $x = (x_1, x_2)$ corresponding to the importance of the second- and third-order information, we solve the following optimization problem so that the observed edges of the new network $W_{new}$ have the highest probability of occurrence and the unobserved edges have the lowest probability of occurrence:

$$\max_x \prod_{I_{ij}^E = 1} M_{ij}(x) \prod_{I_{ij}^N = 1} (1 - M_{ij}(x)) \quad (13)$$

$$s.t. \begin{cases} \sum_i x_i = 1, \\ x_i \geq 0 \end{cases} \quad (14)$$

To facilitate the calculation, we transform the objective function into its log-likelihood value:

$$\max_x \sum_{I_{ij}^E = 1} \log M_{ij}(x) + \sum_{I_{ij}^N = 1} \log(1 - M_{ij}(x)) \quad (15)$$

$$s.t. \begin{cases} \sum_i x_i = 1, \\ x_i \geq 0 \end{cases} \quad (16)$$

We use the Sequential Least Squares Programming (SLSQP) [12] algorithm to solve this optimization problem, which uses the Han-Powell quasi-Newton method with a BFGS update and an L1-test function. SLSQP can be executed directly by the *optimize.minimize* function of the python package *scipy*.

It is worth noting that using another two auxiliary networks from the same temporal system to aid in the search for the optimal parameter is based on the underlying assumption that the emergence of temporal network edges conforms to an invariant pattern, so that we can first explore how the network utilizes local information to connect edges, and then let the results better guide link prediction.

**D. Framework of MOLI**

In conclusion, there are two steps in using the framework MOLI to predict possible edges for temporal networks (Fig. 1(b)):

**Step 1 (Learning the order coefficient $x$):**

Given $W_{old}$, $W_{new}$, initial value $x_0$ and tolerance $eps$ as input, we derive the optimal order coefficient $x^{opt}$ based on the following steps:

● Compute $I^E, I^N, I$:

$$I^E(i,j) = \begin{cases} 1 & W_{old}(i,j) = 0 \text{ \& } W_{new}(i,j) \neq 0 \\ 0 & else \end{cases}$$

$$I^N(i,j) = \begin{cases} 1 & W_{old}(i,j) = 0 \text{ \& } W_{new}(i,j) = 0 \\ 0 & else \end{cases}$$

$$I(i,j) = \begin{cases} 1 & W_{old}(i,j) = 0 \\ 0 & else \end{cases}$$

● Define $M^0(x) = F(W_{old}, x) * I$;
● Define $M(x) = M^0(x) / \sum_{i,j} M_{ij}^0(x)$;
● Solve the optimal solution $x^{opt}$ of equation (15) and (16) using the SLSQP algorithm of the python package *scipy*.

**Step 2 (Link prediction):**

Using the order coefficient $x^{opt} = (x_1^{opt}, x_2^{opt})$ obtained in step 1 as input, we now predict missing links of the observed network $G_{obs}$ based on equation (9). The confidence score matrix of potential edges between two unconnected nodes can be calculated by:

$$G_{MOLI} = D(x_1^{opt} P_{obs}^2 + x_2^{opt} P_{obs}^3) * I_{obs} \qquad (17)$$

where $I_{obs}$ marks those missing edges of the observed network:

$$I_{obs}(i,j) = \begin{cases} 1 & W_{obs}(i,j) = 0 \\ 0 & else \end{cases} \qquad (18)$$

By calculating the AUROC and AUPR scores with the ground truth network $G_{pre}$, we can evaluate the link prediction effectiveness of the MOLI method.

### III. Comparative methods

Numerous methods have been developed for link prediction, most of them are similarity-based methods that utilize the information of common neighbors in different ways, such as Jaccard Similarity [14] and Adamic and Adar's index [17]. There are also some algorithms built based on the global network structure, such as Katz index [19], and Rooted PageRank algorithm [21]. Now we introduce those comparative link prediction methods briefly that we will compare with in the next section.

**1. Common Neighbors** (CN) [13] directly uses the number of common neighbors between two unconnected nodes as the edge confidence score and can be computed based on the adjacency matrix $A$:

$$CN(i,j) = (A^2)_{ij} \qquad (19)$$

**2. Jaccard Similarity** (JS) [14] is the number of common neighbors of two unconnected nodes divided by the number of all distinct neighbors of the two nodes:

$$JS(i,j) = \frac{|N_i \cap N_j|}{|N_i \cup N_j|} \qquad (20)$$

where $N_i$ represents the set of all neighbors of node $i$.

**3. Degree Product** (DP) [15], also known as Preferential Attachment, is applied to generate a growing scale-free network, and can be calculated by multiplying the degrees of two unconnected nodes:

$$DP(i,j) = d_i d_j \qquad (21)$$

**4. Length Three** (L3) [5] is calculated based on the third-order adjacency matrix and normalized by the degree of distant nodes of the path:

$$L3(i,j) = \sum_{k \neq l} \frac{A_{ik} A_{kl} A_{lj}}{\sqrt{d_k d_l}} \qquad (22)$$

**5. Resource Allocation** (RA) [16] is the sum of the reciprocal of the degree of common neighbors of two unconnected nodes:

$$RA(i,j) = \sum_{k \in N_i \cap N_j} \frac{1}{d_k} \qquad (23)$$

Notice that when the input network is unweighted, the RA algorithm is equivalent to the second-order diffused weight matrix of our MOLI algorithm because the equation (23) can be rewritten into:

$$RA(i,j) = \sum_{k \in N_i \cap N_j} \frac{1}{N_k} = \sum_k \left( \frac{W_{ik} W_{kj}}{\sum_m W_{km}} \right)$$

For MOLI, the second-order diffused weight matrix $W^{(2)}$ which computes based on the second-order transition matrix $P^{(2)}$ can be expressed as follows:

$$P_{ij} = \frac{W_{ij}}{\sum_m W_{im}}$$

$$P_{ij}^{(2)} = \sum_k P_{ik} P_{kj} = \sum_k \left( \frac{W_{ik}}{\sum_m W_{im}} \frac{W_{kj}}{\sum_m W_{km}} \right)$$

$$= \frac{1}{\sum_m W_{im}} \sum_k \left( \frac{W_{ik} W_{kj}}{\sum_m W_{km}} \right)$$

$$W_{ij}^{(2)} = d_i P_{ij}^{(2)} = \frac{\sum_m W_{im}}{\sum_m W_{im}} \sum_k \left( \frac{W_{ik} W_{kj}}{\sum_m W_{km}} \right) = \sum_k \left( \frac{W_{ik} W_{kj}}{\sum_m W_{km}} \right)$$

So, in later experiments, we found that sometimes the RA algorithm can get as good results as our MOLI algorithm when the network is second-order information dominated.

**6. Adamic and Adar's index** (AA) [17] is very similar to RA and can be computed by summing of the reciprocal of the logarithm of the degree of common neighbors of two unconnected nodes:

$$AA(i,j) = \sum_{k \in N_i \cap N_j} \frac{1}{\log(d_k)} \qquad (24)$$

**7. Association Strength** (AS) [18], also known as Leicht–Holme–Newman Local Index (LHNL), predicts edge confidence scores by the association strength of neighbors, and can be obtained by calculating the ratio of the number of paths of length two between two nodes to the expected number of paths of the same length between them [1]:

$$AS(i,j) = \frac{|N_i \cap N_j|}{|N_i| \cdot |N_j|} \qquad (25)$$

**8. Katz Index** (Katz) [19] aggregates network information over all the paths between two nodes and gives more attention to the shorter paths:

$$Katz(A) = \beta A + \beta^2 A^2 + \cdots = (I - \beta A)^{-1} - I \qquad (26)$$

Here $\beta$ is the parameter controlling the path weights of different lengths, and to ensure the convergence of this infinite series, $\beta$ should satisfy $\beta < 1/\lambda_1$, where $\lambda_1$ is the maximum eigenvalue of the adjacency matrix $A$ [1].

**9. SimRank** (SR) [20] is developed based on the network structure

with the assumption that two objects are similar if they are related to similar objects:

$$SR(i,j) = \frac{C}{|N(i)||N(j)|} \sum_{k=1}^{|N(i)|} \sum_{l=1}^{|N(j)|} SR(N_k(i), N_l(j)) \quad (27)$$

That is, the similarity between nodes $i$ and $j$ is the average similarity between the neighbors of $i$ and neighbors of $j$. This formula allows us to recursively calculate the similarity between two nodes and diffusely use the information of the whole network.

**10. Rooted PageRank** (RPR) [21] uses the concept of PageRank [22] incorporated with a rooted random walk. The similarity between two nodes can be measured by the probability in a random walk where the walker moves to an arbitrary neighboring node with stationary probability $\alpha$ and returns to itself with probability $1 - \alpha$. Mathematically, the calculation formula is as follows:

$$RPR = (1-\alpha)(1-\alpha\widehat{N})^{-1} \quad (28)$$

where $\widehat{N}$ is the normalized adjacency matrix.

**11. Network Refinement** (NR-F) [6] takes a noisy network as input and outputs a network on the same vertex set with adjusted edge weights. NR-F is a network diffusion method that we developed previously, and MOLI can actually be seen as its local version with variable order coefficient. NR-F is also defined based on three operators: $g$, $f_m$ and $h$. The operators $g$ and $h$ are the same with MOLI in equations (4) and (5), which help us map the random walk diffusion into the underlying graph. The operator $f_m$ transforms a transition matrix $P$ to another by adding the probability of all paths of different length $k$ joining two nodes, with a smaller weight coefficient $1/m^k$ for a longer path of length $k$:

$$f_m(P) = \frac{\sum_k (P^k/m^k)}{\sum_k (1/m^k)} = (m-1)P(mI-P)^{-1} \quad (29)$$

For the NR-F model, to map the diffusion process of random walk on the graph defined by $f_m$ onto the diffusion process on the graph, we wrap the operator $f_m$ in operators $g$ and $h$ to get the composite operator:

$$NR\_F(W) = h(f_m(g(W))) \quad (30)$$

In conclusion, CN, JS, AA, AS, and RA algorithms are all local methods that use only the second-order network information; Katz, SR, RPR, and NR-F methods are all global methods that use multi-order network information. In addition, we classify L3 and DP algorithms as other methods. In the following, we compare these 11 algorithms with MOLI for link prediction on 11 datasets.

### IV. Results

In this section, we demonstrate the effectiveness of our MOLI method in solving the link prediction problem through experiments on four simulated temporal networks (based on BA network [23], ER network [24], Les Misérables Character Network [25], and Rhesus cerebral cortex network [26]) and seven real-world temporal networks (Online social network [27]; European email network [8]; Drug-Drug Interaction network [9]; Protein-Protein Interaction network [7]; MOOC online course network [28]; Wiki-talk network [8]; Reddit hyperlinks network [29]) that cover various domains including biological networks, social networks, and random networks, etc.

Notice that the network evolution problem we discussed here involves only the growth of edges without considering the nodes. Therefore, we set the temporal networks to the same number of nodes for each experiment.

#### A. Simulated networks

We generated the simulated networks as follows: we used four undirected, unweighted networks as the base networks (details are given in Fig. 3(a)-(d)) with the network properties shown in Table 1, and randomly added 10% of the edges to the base networks according to the weight matrix obtained from the MOLI model whose order coefficient was set to two cases: [0,1] and [1,0]. These two settings represent two different network patterns: dominated by second-order information or third-order information. Collectively, we have a total of eight simulated network settings.

Under each configuration, we generated 100 simulated networks and separated them into two groups. Thirty of these networks were used to learn the order coefficient, and their average solution was compared with the preset order coefficient to verify that the model found the network formation pattern. The remaining 70 networks are used to test the performance of link prediction of MOLI and compare the results with the other 11 methods, as shown in Fig. 3(e).

Table 2 and Table 3 show the link prediction results for the simulated networks we generated on the BA network with order coefficients set to [1,0] and [0,1], respectively. The optimal order coefficients obtained by solving the optimization problem under the two settings are [1, 0] and [0.027, 0.973], which confirms that if the networks are generated according to the pattern we have assumed, we can obtain them by solving the optimization problem in equation (15) and (16). The results of the other three simulated networks are similar, which are given in Supplementary Tables S1-S6. It is unsurprising to see that when we set the network to be dominated by second-order information (Table 2), the common neighbors-based approaches perform well while L3 performs poorly;

| Network | #Nodes | #Edges | Average degree | Diameter | Clustering coefficient |
|---|---|---|---|---|---|
| BA | 50 | 141 | 5.64 | 4 | 0.19316 |
| ER | 50 | 383 | 15.32 | 3 | 0.30883 |
| Les Misérables Character | 77 | 254 | 6.60 | 5 | 0.57314 |
| Rhesus cerebral cortex | 91 | 1401 | 30.79 | 3 | 0.74240 |

Table 1. The network properties of the four base networks for simulation experiments.

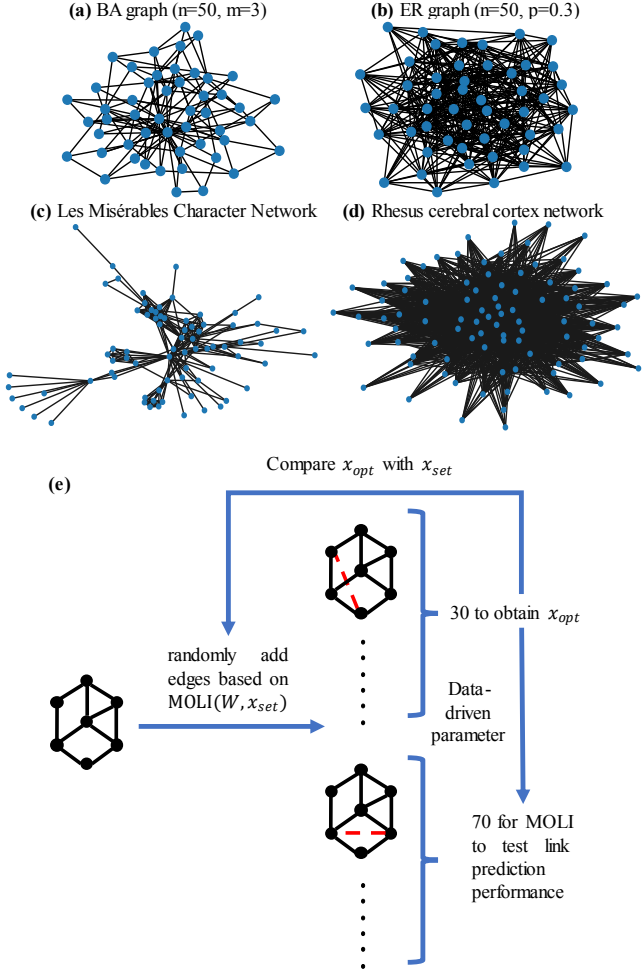

FIG. 3. (a)-(d) Four base networks for generating simulated networks. (a) The BA graph with $n=50$ (number of nodes of the graph) and $m=3$ (number of nodes added each time). (b) The ER graph with $n=50$ (number of nodes of the graph) and $p=0.3$ (connection probability). (c) The Les Misérables Character network describes the relationships between people of the book, and can be downloaded from http://www-personal.umich.edu/~mejn/netdata/. (d) The rhesus cerebral cortex network is the mapping of axonal tracts between neurons, and can be obtained from https://neurodata.io/project/connectomes/. (e) Schematic of how the simulated network is generated.

conversely, when we set the network to be dominated by third-order information (Table 3), L3 achieves superior results while the common neighbors-based approaches give poorer results. In both cases, MOLI not only finds the optimal order coefficients, but also obtains almost the best link prediction performance. It proves that MOLI can make better use of network information and that we should investigate the network edge formation pattern before predicting potential links.

| methods | AUROC | AUPR |
|---|---|---|
| **Local methods (second-order)** | | |
| CN | 0.83476 ± 0.00116 | 0.08128 ± 0.00159 |
| JS | 0.80115 ± 0.00109 | 0.04561 ± 0.00044 |
| AA | 0.85417 ± 0.00097 | **0.08966 ± 0.00180** |
| AS | 0.73404 ± 0.00135 | 0.02920 ± 0.00021 |
| RA | **0.85479 ± 0.00093** | 0.08667 ± 0.00173 |
| **Global methods** | | |
| Katz | 0.83059 ± 0.00173 | 0.08173 ± 0.00159 |
| SR | 0.75077 ± 0.00128 | 0.03256 ± 0.00025 |
| RPR | 0.82326 ± 0.00203 | 0.08648 ± 0.00234 |
| NR-F | 0.84493 ± 0.00134 | 0.08890 ± 0.00241 |
| **Other methods** | | |
| DP | 0.72616 ± 0.00457 | 0.05910 ± 0.00206 |
| L3 (third-order) | 0.64287 ± 0.00598 | 0.05401 ± 0.00224 |
| MOLI | **0.85513 ± 0.00089** | **0.08685 ± 0.00172** |

Table 2. AUROC and AUPR scores of different link prediction methods on simulated BA network under the setting of the second-order information dominated, namely, the order coefficient we have set is [1,0]. The top 2 results in terms of AUROC and AUPR are marked in bold separately.

| methods | AUROC | AUPR |
|---|---|---|
| **Local methods (second-order)** | | |
| CN | 0.58523 ± 0.00401 | 0.02917 ± 0.00040 |
| JS | 0.53315 ± 0.00296 | 0.01532 ± 0.00001 |
| AA | 0.58134 ± 0.00389 | 0.02958 ± 0.00035 |
| AS | 0.49810 ± 0.00226 | 0.01308 ± 0.00000 |
| RA | 0.57974 ± 0.00384 | 0.02816 ± 0.00031 |
| **Global methods** | | |
| Katz | 0.64549 ± 0.00415 | 0.03591 ± 0.00049 |
| SR | 0.48516 ± 0.00313 | 0.01295 ± 0.00000 |
| RPR | 0.67109 ± 0.00454 | 0.03781 ± 0.00054 |
| NR-F | 0.65606 ± 0.00403 | 0.03771 ± 0.00054 |
| **Other methods** | | |
| DP | 0.68917 ± 0.00533 | 0.04211 ± 0.00072 |
| L3 (third-order) | **0.72757 ± 0.00390** | **0.04702 ± 0.00073** |
| MOLI | **0.73193 ± 0.00407** | **0.04754 ± 0.00071** |

Table 3. AUROC and AUPR scores of different link prediction methods on simulated BA network under the setting of the third-order information dominated, namely, the order coefficient we have set is [0,1]. The top 2 results in terms of AUROC and AUPR are marked in bold separately.

**B. Real-world temporal networks**

**A. Online Social networks**

This dataset is comprised of private messages sent on a Facebook-like online social network at the University of California [27] and can be download from https://snap.stanford.edu/data/CollegeMsg.html. Online users can initiate conversations with anyone of interest based on the profile information. When user $i$ sends a private message to user $j$, there will be an edge connects them. We took the records of 193 days and divided them into three groups (t=1; t=2; t=3) in chronological order, then constructed three temporal social networks and took out the largest connected component. We use the network pair at t=1 and t=2 to learn the order coefficient, and the network pair at t=2 and t=3 to test the link prediction performance of various methods. The properties of the three temporal networks are shown in Table 4.

The optimal coefficient we got from MOLI, surprisingly, is [0.03, 0.97], which implies that online social network is dominated by the third-order information. That contradicts our empirical experience that social networks connect their edges based on common neighbors. A more insightful interpretation will help explain its plausibility: because of the privateness of online socialization, even if $i$ and $j$ have friends in common, the two may not know each other. But if both $i$ and $j$ are connected to $k$, then it is possible that they are interested in similar types of people. So, if $i$ checks the profile of $l$ and initiates a private chat, then $j$ is likely to be interested in $l$ as well. We call it the Quadrilateral Closure Principle (QCP), which preferentially connects paths of length three (as illustrated of the green edge in Fig. 1(a)), in contrast to the Triadic Closure Principle (TCP). The results shown in Table 5 also confirm this discovery, since all the second-order local methods perform unsatisfactorily, while L3 and MOLI get the best link prediction performance. In addition, we point out that even using only the prior second-order information for link prediction, MOLI outperformed other common neighbors-based methods, which shows the superiority of our method in making better use of network information.

| Networks | #Nodes | #Edges | Diameter | Average degree | Maximum degree | Clustering coefficient |
|---|---|---|---|---|---|---|
| Social (t=1) | 1022 | 5334 | 7 | 10.44 | 212 | 0.10519 |
| Social (t=2) | 1022 | 7082 | 7 | 13.86 | 213 | 0.11789 |
| Social (t=3) | 1022 | 8020 | 6 | 15.69 | 217 | 0.12997 |
| Email (t=1) | 837 | 8896 | 7 | 21.26 | 215 | 0.35751 |
| Email (t=2) | 837 | 12373 | 7 | 29.57 | 297 | 0.39410 |
| Email (t=3) | 837 | 14849 | 7 | 35.48 | 337 | 0.41596 |
| DDI (train) | 4267 | 1067911 | 5 | 500.54 | 2234 | 0.51426 |
| DDI (validation) | 4267 | 1201400 | 5 | 563.11 | 2477 | 0.57545 |
| DDI (test) | 4267 | 1334889 | 5 | 625.68 | 2477 | 0.63853 |
| PPI (2011) | 4923 | 58586 | 6 | 23.80 | 2323 | 0.25811 |
| PPI (2013) | 4923 | 75879 | 5 | 30.83 | 2924 | 0.30660 |
| PPI (2018) | 4923 | 91661 | 5 | 37.24 | 2705 | 0.32114 |
| MOOC (t=1) | 5171 | 87478 | 4 | 33.83 | 4885 | 0 |
| MOOC (t=2) | 5171 | 94696 | 4 | 36.63 | 4889 | 0 |

Table 4. The network properties of the five real-world temporal networks shown in the main text.

The link prediction results on the temporal Wiki-talk networks [8] and Reddit hyperlinks networks [29], whose network properties are shown in Supplementary Table S7, are similar to those of online social networks, and we give the results in Supplementary Tables S8 and S9, respectively, without going into detail here.

| methods | AUROC | AUPR |
|---|---|---|
| **Local methods (second-order)** | | |
| CN | 0. 688540 | 0. 005768 |
| JS | 0. 649706 | 0. 002992 |
| AA | 0. 695782 | 0. 006100 |
| AS | 0. 625232 | 0. 002615 |
| RA | 0. 697989 | 0. 006419 |
| **Global methods** | | |
| Katz | 0. 804893 | 0. 008125 |
| SR | 0. 535892 | 0. 001869 |
| RPR | 0. 817029 | 0. 007296 |
| NR-F | 0. 830882 | 0. 009763 |
| **Other methods** | | |
| DP | 0. 837927 | 0. 011002 |
| L3 (third-order) | **0. 857701** | **0. 015488** |
| MOLI | **0. 859212** | **0. 015322** |
| MOLI([0.5, 0.5]) | 0. 838043 | 0. 010902 |
| MOLI([1, 0]) | 0. 697857 | 0. 006415 |

Table 5. Link prediction results of Facebook-like online social networks. The top 2 results in terms of AUROC and AUPR are marked in bold separately.

B. European email network

The European email temporal networks describe the emails sent to each other by people at a large European research institution over an 18-month period from October 2003 to May 2005 (http://snap.stanford.edu/data/email-Eu-core-temporal.html) [8]. Given a set of email messages, each node of the email network corresponds to an email address and each edge represents an email. We create an undirected edge between nodes $i$ and $j$, if $i$ sent at least one message to $j$. The email records over 18 months were divided into three groups in chronological order and used to construct three temporal networks. The properties of those networks are shown in Table 4.

The optimal order coefficient output by MOLI is $[0.48, 0.52]$, which indicates that the second-order and the third-order information are equally important in predicting the edges for the European email temporal networks, and it is not sufficient to consider only one of them. This is verified in the results of Table 6, where MOLI is more effective than algorithms using only the second- or third-order neighbors. Furthermore, same as before, MOLI based simply on second-order information outperforms other second-order methods, and MOLI based solely on third-order information outperforms L3, indicating that MOLI uses network information more efficiently and rationally than other link prediction methods. In addition, we can also see that several global approaches do not perform well, which suggests that local information is effective enough in social networks, while higher-order information is not reliable.

| methods | AUROC | AUPR |
|---|---|---|
| **Local methods (second-order)** | | |
| CN | 0.89412 | 0.08918 |
| JS | 0.87479 | 0.07343 |
| AA | 0.89937 | 0.09401 |
| AS | 0.79944 | 0.01998 |
| RA | 0.90165 | **0.09578** |
| **Global methods** | | |
| Katz | 0.88813 | 0.07452 |
| SR | 0.80639 | 0.02206 |
| RPR | 0.89945 | 0.06356 |
| NR-F | **0.91092** | 0.09287 |
| **Other methods** | | |
| DP | 0.82632 | 0.04915 |
| L3 (third-order) | 0.89741 | 0.08196 |
| MOLI | **0.91323** | **0.09606** |
| MOLI([1,0]) | 0.90166 | **0.09578** |
| MOLI([0,1]) | 0.90788 | 0.08897 |

Table 6. Link prediction results on European email temporal networks. The top 2 results in terms of AUROC and AUPR are marked in bold separately.

Although the results given by MOLI were validated, it's also somewhat different from our expectations. Because email communication networks are a type of social network, and therefore TCP is applicable, resulting in a network that should be connected in a second-order information dominant manner, i.e., preferentially connecting nodes that have common neighbors. But this is not exactly the case, because third-order information leads to some edge connections as well. For example, if A and B have friends in common, which might indicate that they belong to the same social group, then if A and C have email correspondence, B has likely sent emails with C as well. In summary, both the second-order and the third-order neighbors are interpretable in email communication networks.

C.  Drug-Drug Interaction network

The Drug-Drug Interaction (DDI) dataset can be found at the link property prediction task of the *Open Graph Benchmark (OGB)* competition from https://ogb.stanford.edu/docs/linkprop/#ogbl-ddi. Each node of the DDI network represents an FDA-approved or experimental drug, while the edges reflect drug interactions in which the combined effect of two drugs differs significantly from the anticipated result when the drugs act separately. OGB splits the dataset into three according to what proteins those drugs target. This indicates that drugs in the test set have a different biological mechanism of action and work differently in the body than those in the train and validation sets [9][10]. The properties of the three networks (train, validation, test) are given in Table 4.

We treat the train network as the old network and the validation network as the new network to explore the pattern of edge emergence, and obtain the optimal order coefficient of $[0.99, 0.01]$, which indicates that we should use the second-order information to predict missing edges. That is consistent with the empirical fact that if drugs A and B interact with each other, and B and C interact with each other, there is a high probability that jointly using drugs A and C will also affect each other's efficacy. The results shown in Table 7 verified that common neighbors-based algorithms generally outperform L3, and the effectiveness of MOLI is the best among all the 12 methods. And as we previously proved that the RA algorithm is equivalent to the second-order weight matrix of our MOLI algorithm, RA also works relatively well in this experiment.

| methods | AUROC | AUPR |
|---|---|---|
| **Local methods (second-order)** | | |
| CN | 0.95737 | 0.46757 |
| JS | 0.96257 | 0.36966 |
| AA | 0.95911 | 0.47825 |
| AS | 0.84559 | 0.04928 |
| RA | **0.96762** | **0.50468** |
| **Global methods** | | |
| Katz | 0.94232 | 0.33681 |
| SR | 0.84970 | 0.05325 |
| RPR | 0.94443 | 0.26218 |
| NR-F | 0.95793 | 0.40047 |
| **Other methods** | | |
| DP | 0.88951 | 0.13032 |
| L3 (third-order) | 0.93924 | 0.29142 |
| MOLI | **0.96762** | **0.50410** |
| MOLI([0.5, 0.5]) | 0.96304 | 0.44072 |
| MOLI([0, 1]) | 0.95242 | 0.33268 |

Table 7. Link prediction results on the Drug-Drug Interaction networks. The top 2 results in terms of AUROC and AUPR are marked in bold separately.

### D. Protein-Protein Interaction network

Protein-Protein interaction (PPI) networks have been proven to be third-order information-dominated networks in earlier research [5], and we were interested to know if we could get similar results. We first constructed the *Saccharomyces cerevisiae* PPI networks as in [7]:

(1) We downloaded three snapshots of the *S. cerevisiae* PPI datasets on the BioGRID database from https://downloads.thebiogrid.org/BioGRID/Release-Archive/ whose interactions were verified by at least one wet-lab experiment. These three datasets were curated by BioGRID until 2011, 2013, and 2018, respectively (version 3.1.80, 3.2.106, and 3.4.157).

(2) We chose the 5001 verified ORFs [11] as nodes of the networks and extracted the physical interactions between these nodes in three datasets as edges of the networks. After removing isolated nodes, redundant edges, and the edges in the old version of BioGRID but not present in the new version of BioGRID, we obtained three undirected networks as shown in Table 4. Besides, we followed the work of [11] and created the confidence score for each edge according to the publications supporting them.

After constructing three temporal PPI networks, we got the optimal order coefficient as [0.06, 0.94] by solving the optimization problem of network 2011 and 2013, which confirms that third-order information should be used instead of common neighbors information in PPI networks. Then we used the network 2013 and 2018 to evaluate the performance of link prediction. And we can see from the results in Table 8 that MOLI outperforms L3 and certainly outperforms other second-order local methods.

| methods | AUROC | AUPR |
|---|---|---|
| **Local methods (second-order)** | | |
| CN | 0.74023 | 0.00988 |
| JS | 0.58518 | 0.00221 |
| AA | 0.74476 | 0.00994 |
| AS | 0.53068 | 0.00184 |
| RA | 0.73973 | 0.00881 |
| **Global methods** | | |
| Katz | 0.77609 | 0.01062 |
| SR | 0.51312 | 0.00178 |
| RPR | **0.82253** | 0.01086 |
| NR-F | 0.81387 | 0.01118 |
| **Other methods** | | |
| DP | 0.79742 | 0.00915 |
| L3 (third-order) | 0.80662 | **0.01123** |
| MOLI | **0.81569** | **0.01203** |
| MOLI([0.5, 0.5]) | 0.81010 | 0.01140 |
| MOLI([1, 0]) | 0.75422 | 0.00965 |

Table 8. Link prediction results on temporal Protein-Protein Interaction networks. The top 2 results in terms of AUROC and AUPR are marked in bold separately.

### E. MOOC online course network

Finally, we show in this experiment that we can also get the parameter of MOLI based on prior knowledge, without solving the optimization problem, when there are no two other auxiliary temporal networks for us to explore the network formation pattern, or, prior knowledge is reliable enough.

Massive Open Online Courses (MOOC) are open-access online courses that anybody can take for free. The dataset, compiled by Srijan *et al.* [28], consists of 7047 users interacting with 98 items (videos, answers, etc.), generating over 411,749 actions (watching videos, answering questions, etc.). We constructed two temporal datasets in the same way as before, and the properties of these two networks are also listed in Table 4. Note that the networks are constructed as bipartite networks, so common neighbors are useless in predicting edges. We, therefore, can directly set the parameter of MOLI to [0,1], i.e., only third-order information is used for link prediction. The results are shown in Table 9, and we can see that MOLI still gives the best performance, even better than all the global methods. Moreover, even if there is no prior knowledge, MOLI with the equal order coefficient [0.5, 0.5] can always give relatively good prediction, which is also shown in previous experiments.

| methods | AUROC | AUPR |
|---|---|---|
| **Local methods (second-order)** | | |
| CN | 0.005872 | 0.000271 |
| JS | 0.005872 | 0.000271 |
| AA | 0.005872 | 0.000271 |
| AS | 0.005872 | 0.000271 |
| RA | 0.005872 | 0.000271 |
| **Global methods** | | |
| Katz | 0.913282 | 0.073851 |
| SR | 0.003397 | 0.000272 |
| RPR | 0.994018 | 0.074164 |
| NR-F | 0.983692 | 0.115529 |
| **Other methods** |  |  |
| DP | 0.975125 | 0.097545 |
| L3 (third-order) | **0.997455** | **0.097658** |
| MOLI | **0.997746** | **0.134669** |
| MOLI([0.5, 0.5]) | 0.991705 | 0.119855 |
| MOLI([1, 0]) | 0.005896 | 0.000271 |

Table 9. Link prediction results on temporal MOOC online course networks. The top 2 results in terms of AUROC and AUPR are marked in bold separately.

In addition, we point out that we can use the clustering coefficient of the network to roughly estimate whether the network will add edges based on the common neighbors. This is because the network clustering coefficient reflects the connectedness of triadic closure, if the clustering coefficient is 0, then the network is a bipartite graph, and of course, the second-order information fails to predict edges; if the clustering coefficient is relatively large, then it is likely that the network is connected by common neighbors. This can be corroborated by comparing the network clustering coefficients in Table 4 with the results of each experiment.

## V. Conclusion

In this paper, we proposed MOLI, a novel link prediction method that can explore network formation pattern for better predicting possible edges. The parameter of MOLI can be either data-driven by solving an optimization problem, or prior-driven based on reliable experience.

We not only prove the correctness and superiority of MOLI on four simulated networks, but also get the conclusion that different networks have different patterns of edge emergence on seven real-world temporal networks. Some conclusions given by MOLI are conflict with the empirical experience. For example, we revealed that the second- and third-order local information are equally important in predicting edges of communication networks, while some online social networks that empirically considered obeying TCP actually confirm the Quadrilateral Closure Principle, which preferentially connects paths of length three. That shows prior-driven algorithms can sometimes be highly biased, we should explore the network development pattern adaptively before predicting edges.

We don't use higher-order information for link prediction task, since it not only increases the computational complexity, but also untrustworthy. As stated before, MOLI is a local version of NR-F, which considers only the diffusion of second- and third-order network information. The results indicate that NR-F does not work efficiently, although it outperforms other global methods and some local methods in most experiments. That is because, when we generally use the n-th power of the adjacency matrix (or transition probability matrix) to characterize the n-th order information of the network, there will be inevitably redundancy of information in the higher-order matrix due to the duplicate nodes on the long paths. Besides, neighborhoods information is indeed more reliable than higher-order information for most networks. Therefore, MOLI can efficiently and effectively achieve better performance by only using the second- and third-order local information.

In summary, MOLI, as a better link prediction method that fits the development pattern of a specific network, can corroborate with the empirical experience, and gradually improve our recognition of various network development patterns.

### Acknowledgements

This work has been supported by the National Key Research and Development Program of China (No. 2020YFA0712402) and the National Natural Science Foundation of China (No. 11631014).

### Competing interests

The authors declare no competing interests.

### Code available

MOLI can be downloaded at https://github.com/Wu-Lab/MOLI.

# Supplementary Tables S1-S9

| methods | AUROC | AUPR |
|---|---|---|
| **Local methods (second-order)** | | |
| Common Neighbors | 0.62716 ± 0.00185 | 0.07447 ± 0.00024 |
| Jaccard Similarity | 0.62310 ± 0.00197 | 0.07426 ± 0.00025 |
| Adamic Adar | 0.63189 ± 0.00190 | 0.07647 ± 0.00025 |
| Association Strength | 0.59480 ± 0.00206 | 0.06418 ± 0.00016 |
| Resource Allocation | **0.63291 ± 0.00192** | **0.07682 ± 0.00026** |
| **Global methods** | | |
| Katz | 0.62587 ± 0.00198 | 0.07536 ± 0.00023 |
| SimRank | 0.59717 ± 0.00205 | 0.06490 ± 0.00016 |
| Rooted PageRank | 0.61637 ± 0.00215 | 0.07248 ± 0.00019 |
| NR-F | 0.62553 ± 0.00209 | 0.07494 ± 0.00022 |
| **Other methods** | | |
| Degree Product | 0.58802 ± 0.00224 | 0.06575 ± 0.00012 |
| L3 (third-order) | 0.57647 ± 0.00267 | 0.06373 ± 0.00017 |
| MOLI | **0.63261 ± 0.00197** | **0.07674 ± 0.00026** |

**Table S1.** AUROC and AUPR scores of different link prediction methods on simulated **ER** networks under the setting of the **second**-order information dominated. The optimal coefficient given by the optimization problem is [0.9, 0.1]. The top 2 results in terms of AUROC and AUPR are marked in bold separately.

| methods | AUROC | AUPR |
|---|---|---|
| **Local methods (second-order)** | | |
| Common Neighbors | 0.57251 ± 0.00223 | 0.06392 ± 0.00028 |
| Jaccard Similarity | 0.54699 ± 0.00216 | 0.05783 ± 0.00017 |
| Adamic Adar | 0.57079 ± 0.00211 | 0.06450 ± 0.00027 |
| Association Strength | 0.49948 ± 0.00189 | 0.04611 ± 0.00004 |
| Resource Allocation | 0.56990 ± 0.00200 | 0.06376 ± 0.00026 |
| **Global methods** | | |
| Katz | 0.58339 ± 0.00226 | 0.06695 ± 0.00030 |
| SimRank | 0.50222 ± 0.00191 | 0.04664 ± 0.00004 |
| Rooted PageRank | 0.60162 ± 0.00202 | 0.07051 ± 0.00035 |
| NR-F | 0.59416 ± 0.00197 | 0.06876 ± 0.00031 |
| **Other methods** | | |
| Degree Product | 0.61101 ± 0.00177 | 0.07037 ± 0.00032 |
| L3 (third-order) | **0.61131 ± 0.00185** | **0.07177 ± 0.00037** |
| MOLI | **0.60880 ± 0.00188** | **0.07180 ± 0.00037** |

**Table S2.** AUROC and AUPR scores of different link prediction methods on simulated **ER** networks under the setting of the **third**-order information dominated. The optimal coefficient given by the optimization problem is [0.22, 0.78]. The top 2 results in terms of AUROC and AUPR are marked in bold separately.

| methods | AUROC | AUPR |
|---|---|---|
| **Local methods (second-order)** | | |
| Common Neighbors | 0.66659 ± 0.00052 | 0.13311 ± 0.00052 |
| Jaccard Similarity | 0.59162 ± 0.00049 | 0.07111 ± 0.00005 |
| Adamic Adar | 0.66788 ± 0.00052 | 0.13385 ± 0.00051 |
| Association Strength | 0.46693 ± 0.00055 | 0.04933 ± 0.00002 |
| Resource Allocation | **0.66867 ± 0.00050** | **0.13425 ± 0.00050** |
| **Global methods** | | |
| Katz | 0.66130 ± 0.00052 | 0.13089 ± 0.00052 |
| SimRank | 0.48900 ± 0.00050 | 0.05056 ± 0.00002 |
| Rooted PageRank | 0.62518 ± 0.00043 | 0.11912 ± 0.00050 |
| NR-F | 0.65022 ± 0.00049 | 0.12766 ± 0.00050 |
| **Other methods** | | |
| Degree Product | 0.62853 ± 0.00049 | 0.12116 ± 0.00050 |
| L3 (third-order) | 0.62484 ± 0.00047 | 0.11798 ± 0.00047 |
| MOLI | **0.66859 ± 0.00050** | **0.13421 ± 0.00050** |

**Table S3.** AUROC and AUPR scores of different link prediction methods on the simulated **rhesus cerebral cortex** network under the setting of the **second**-order information dominated. The optimal coefficient given by the optimization problem is [0.97, 0.03]. The top 2 results in terms of AUROC and AUPR are marked in bold separately.

| methods | AUROC | AUPR |
| --- | --- | --- |
| **Local methods (second-order)** | | |
| Common Neighbors | 0.64302 ± 0.00052 | 0.11701 ± 0.00033 |
| Jaccard Similarity | 0.41004 ± 0.00053 | 0.04379 ± 0.00002 |
| Adamic Adar | 0.64384 ± 0.00051 | 0.11813 ± 0.00034 |
| Association Strength | 0.28036 ± 0.00050 | 0.03358 ± 0.00000 |
| Resource Allocation | 0.64413 ± 0.00051 | 0.11949 ± 0.00031 |
| **Global methods** | | |
| Katz | 0.70542 ± 0.00053 | 0.14492 ± 0.00044 |
| SimRank | 0.29430 ± 0.00053 | 0.03422 ± 0.00000 |
| Rooted PageRank | 0.76057 ± 0.00049 | 0.16553 ± 0.00046 |
| NR-F | 0.74341 ± 0.00052 | 0.16263 ± 0.00045 |
| **Other methods** | | |
| Degree Product | 0.76309 ± 0.00049 | 0.17013 ± 0.00048 |
| L3 (third-order) | **0.76555 ± 0.00049** | **0.17188 ± 0.00052** |
| MOLI | **0.76609 ± 0.00048** | **0.17182 ± 0.00053** |

**Table S4.** AUROC and AUPR scores of different link prediction methods on simulated **rhesus cerebral cortex** network under the setting of the **third**-order information dominated. The optimal coefficient given by the optimization problem is [0.03, 0.97]. The top 2 results in terms of AUROC and AUPR are marked in bold separately.

| methods | AUROC | AUPR |
| --- | --- | --- |
| **Local methods (second-order)** | | |
| Common Neighbors | 0.88384 ± 0.00010 | 0.22660 ± 0.00119 |
| Jaccard Similarity | 0.82371 ± 0.00012 | 0.08481 ± 0.00008 |
| Adamic Adar | 0.90069 ± 0.00012 | 0.22490 ± 0.00109 |
| Association Strength | 0.78347 ± 0.00007 | 0.06407 ± 0.00003 |
| Resource Allocation | 0.89959 ± 0.00012 | 0.22174 ± 0.00112 |
| **Global methods** | | |
| Katz | 0.89644 ± 0.00014 | 0.21998 ± 0.00115 |
| SimRank | 0.74050 ± 0.00044 | 0.06569 ± 0.00006 |
| Rooted PageRank | 0.91062 ± 0.00010 | 0.21253 ± 0.00080 |
| NR-F | **0.93440 ± 0.00008** | **0.37262 ± 0.00247** |
| **Other methods** | | |
| Degree Product | 0.82433 ± 0.00040 | 0.23184 ± 0.00175 |
| L3 (third-order) | 0.88040 ± 0.00027 | 0.21828 ± 0.00113 |
| MOLI | **0.94242 ± 0.00006** | **0.39115 ± 0.00239** |

**Table S5.** AUROC and AUPR scores of different link prediction methods on simulated **Les Misérables** network under the setting of the **second**-order information dominated. The optimal coefficient given by the optimization problem is [0.98, 0.02]. The top 2 results in terms of AUROC and AUPR are marked in bold separately.

| methods | AUROC | AUPR |
| --- | --- | --- |
| **Local methods (second-order)** | | |
| Common Neighbors | 0.81699 ± 0.00036 | 0.17787 ± 0.00103 |
| Jaccard Similarity | 0.75111 ± 0.00029 | 0.06190 ± 0.00002 |
| Adamic Adar | 0.82591 ± 0.00042 | 0.16460 ± 0.00090 |
| Association Strength | 0.70421 ± 0.00025 | 0.04837 ± 0.00000 |
| Resource Allocation | 0.82380 ± 0.00042 | 0.15461 ± 0.00082 |
| **Global methods** | | |
| Katz | 0.86241 ± 0.00024 | 0.18570 ± 0.00101 |
| SimRank | 0.65539 ± 0.00053 | 0.04301 ± 0.00001 |
| Rooted PageRank | 0.87059 ± 0.00025 | 0.13969 ± 0.00028 |
| NR-F | **0.90108 ± 0.00022** | **0.29734 ± 0.00183** |
| **Other methods** | | |
| Degree Product | 0.85125 ± 0.00039 | 0.23625 ± 0.00145 |
| L3 (third-order) | 0.87293 ± 0.00025 | 0.19721 ± 0.00108 |
| MOLI | **0.90585 ± 0.00020** | **0.30589 ± 0.00189** |

**Table S6.** AUROC and AUPR scores of different link prediction methods on simulated **Les Misérables** network under the setting of the **third**-order information dominated. The optimal coefficient given by the optimization problem is [0.02, 0.98]. The top 2 results in terms of AUROC and AUPR are marked in bold separately.

| Networks | #Nodes | #Edges | Diameter | Average degree | Maximum degree | Clustering coefficient |
|---|---|---|---|---|---|---|
| Wiki-talk (t=1) | 2244 | 8067 | 6 | 7.19 | 1154 | 0.217082 |
| Wiki-talk (t=2) | 2244 | 10627 | 6 | 9.47 | 1201 | 0.245833 |
| Wiki-talk (t=3) | 2244 | 10991 | 6 | 9.80 | 1201 | 0.247693 |
| Reddit (t=1) | 1617 | 2356 | 14 | 2.91 | 139 | 0.059316 |
| Reddit (t=2) | 1617 | 3219 | 12 | 3.98 | 186 | 0.111190 |
| Reddit (t=2) | 1617 | 3934 | 10 | 4.87 | 228 | 0.149360 |

**Table S7.** The network properties of the two real-world temporal networks: Wiki-talk networks and Reddit hyperlinks networks. The corresponding results on the two networks are shown in Supplementary Tables S8 and S9.

| methods | AUROC | AUPR |
|---|---|---|
| **Local methods (second-order)** | | |
| Common Neighbors | 0.833969 | 0.002950 |
| Jaccard Similarity | 0.672962 | 0.000217 |
| Adamic Adar | 0.848407 | 0.003052 |
| Association Strength | 0.611816 | 0.000184 |
| Resource Allocation | 0.846025 | 0.002864 |
| **Global methods** | | |
| Katz | 0.898959 | 0.003427 |
| SimRank | 0.450063 | 0.000119 |
| Rooted PageRank | 0.906007 | 0.002002 |
| NR-F | 0.925442 | 0.003738 |
| **Other methods** | | |
| Degree Product | 0.934907 | 0.003825 |
| L3 (third-order) | **0.937506** | **0.004304** |
| MOLI | **0.939564** | **0.004559** |
| MOLI([0.5, 0.5]) | 0.927267 | 0.003944 |
| MOLI([1, 0]) | 0.850074 | 0.002867 |

**Table S8.** AUROC and AUPR scores of different link prediction methods on Wiki-talk temporal networks. This dataset can be downloaded from http://snap.stanford.edu/data/wiki-talk-temporal.html. The optimal coefficient given by the optimization problem is [0.03, 0.97]. The top 2 results in terms of AUROC and AUPR are marked in bold separately.

| methods | AUROC | AUPR |
| --- | --- | --- |
| **Local methods (second-order)** | | |
| Common Neighbors | 0.762142 | 0.026516 |
| Jaccard Similarity | 0.753548 | 0.004335 |
| Adamic Adar | 0.763565 | 0.028120 |
| Association Strength | 0.750704 | 0.003944 |
| Resource Allocation | 0.763188 | 0.021958 |
| **Global methods** | | |
| Katz | 0.894137 | 0.027745 |
| SimRank | 0.796851 | 0.002047 |
| Rooted PageRank | **0.920314** | 0.011807 |
| NR-F | **0.921772** | 0.031412 |
| **Other methods** | | |
| Degree Product | 0.908016 | 0.032121 |
| L3 (third-order) | 0.869928 | **0.042044** |
| MOLI | 0.882417 | **0.042809** |
| MOLI([0.5, 0.5]) | 0.880547 | 0.032415 |
| MOLI([1, 0]) | 0.763260 | 0.021753 |

**Table S9.** AUROC and AUPR scores of different link prediction methods on Reddit hyperlinks temporal networks. This dataset can be downloaded from http://snap.stanford.edu/data/soc-RedditHyperlinks.html. The optimal coefficient given by the optimization problem is [0.14, 0.86]. The top 2 results in terms of AUROC and AUPR are marked in bold separately.